\documentclass[12pt,twocolumn]{IEEEtran}
\usepackage{fancyhdr}
\fancypagestyle{headings}{%
\cfoot{\thepage}
}
\usepackage{amsmath,amssymb,epsfig,amsthm,psfrag,epsf, multirow,wrapfig, graphicx,dsfont}
\usepackage[english]{babel}
\usepackage{epstopdf}
\usepackage{paralist}
\usepackage[normalem]{ulem}
\usepackage[noadjust]{cite}
\usepackage{subfigure}

\usepackage{cleveref}
\usepackage[linesnumbered,ruled,vlined]{algorithm2e}
\usepackage{booktabs}
\usepackage{extarrows}
  
\usepackage{bm}
\usepackage{caption}

\usepackage{array,multirow}

\title{Artificial Intelligence-Enabled Cellular Networks: \\
A Critical Path to Beyond-5G and 6G}
\author{\IEEEauthorblockN{Rubayet Shafin, Lingjia Liu, Vikram Chandrasekhar, Hao Chen, \\Jeffrey Reed, and Jianzhong (Charlie) Zhang}
\thanks{R. Shafin, L. Liu, and J. Reed are with Virginia Tech, USA. V. Chandrasekhar, H. Chen, and J. Zhang are with Samsung Research America, USA. Supports from National Science Foundation (ECCS-1802710, ECCS-1811497, and CNS-1811720) are acknowledged. The corresponding author is L. Liu (ljliu@ieee.org).}}

\begin{document}
\maketitle
	
\begin{abstract}
Mobile Network Operators (MNOs) are in process of overlaying their conventional macro cellular networks with shorter range cells such as outdoor pico cells. 
The resultant increase in network complexity creates substantial overhead in terms of operating expenses, time, and labor for their planning and management. Artificial intelligence (AI) offers the potential for MNOs to operate their networks in a more organic and cost-efficient manner. We argue that deploying AI in 5G and Beyond will require surmounting significant technical barriers in terms of robustness, performance, and complexity. 
We outline future research directions, identify top 5 challenges, and present a possible roadmap to realize the vision of AI-enabled cellular networks for Beyond-5G and 6G. 
\end{abstract}

\section{Introduction}
Artificial intelligence (AI) is having a transformational effect in every industry and will likely be the foundation of a fourth industrial revolution. Indeed, we are in the middle of the perfect storm propelling AI from advancements in hardware, storage, and software. 
In areas such as computer vision, gaming, and natural language processing, AI has already made significant advancements, 
and their presence is ubiquitous. 
In contrast, the application of AI within the cellular domain, while promising, is still in its nascent stages. Below, we outline key motivations for employing AI-enabled cellular networks.

\begin{asparadesc}
\item[Network Complexity.]
Advancements in error control coding and communication design have resulted in the performance of the point-to-point link being close to the Shannon limit. This has proven effective for designing the 4G LTE-Advanced air interface which (conceptually) consisted of multiple parallel point-to-point links.
However, 5G and future 6G air interfaces will be vastly more complicated due to their complex network topology, multiple numerologies, network coordination schemes, and the diverse nature of end-user applications.
Considering its multifaceted nature, in such complex deployment scenarios, deriving any performance optimum is likely computationally infeasible. AI, however, can tame the network complexity by providing pragmatic, yet competitive performances.

\item[Model Deficit.]
Contemporary cellular systems have been designed with the premise of approximating the end-to-end system behavior using simple modeling approaches that are amenable to clean mathematical analysis. For example, practical systems apply techniques such as digital pre-distortion to linearize the end-to-end model, for which information theory provides a simple closed-form capacity expression. However, in the presence of non-linearities, either due to the underlying wireless channel (e.g. mmWave and Terahertz channels) or device components (e.g. power amplifier), it becomes difficult to analytically model such behaviors in a tractable manner. In contrast, new AI-based detection strategies can be developed to overcome the underlying unknown non-linearities~\cite{mosleh2017brain}. 

\item[Algorithm Deficit.]
There are a variety of scenarios in cellular networks where the optimal algorithms are well characterized, yet are too complex to be implemented in practice. System designers often have to rely on heuristics based on some simple decision making rules. For example, for a point-to-point MIMO link operating with an $M$-ary QAM constellation and $K$ spatial streams, the optimum maximum likelihood receiver incurs prohibitive complexity $O(M^K)$. In practice, most MIMO systems employ linear receivers, e.g. linear minimum mean squared error (MMSE) receiver, which are known to be sub-optimal, yet easy to implement. AI can offer an attractive performance--complexity trade-off in such scenarios. For example, a deep learning based MIMO receiver can provide better performance than linear receivers in a variety of scenarios, while retaining low complexity~\cite{samuel2017deep}. 
\end{asparadesc}

\section{AI for Wireless: Status}
This section overviews the key thrusts in AI for wireless from a fundamental research perspective, and from an industry and standardization perspective. 
Table~\ref{table:PHY_and_Network_Layer} lists key contemporary research works on AI applications relating to cellular networks.

\begin{table*}
\centering
\caption{Overview of research areas on applying AI towards cellular networks.}
\begin{tabular}{ |p{2.6cm}||p{5cm}|p{3cm}|p{4cm}|  }
\hline
\textbf{Layer} & \textbf{Applications}  & \textbf{Learning Method} & \textbf{\centering Tools}\\
\hline \hline
\multirow{5}{*}{\centering PHY \& MAC Layer} & Channel estimation and prediction  & Supervised &   CNN\cite{neumann2018learning}\\
\cline{2-4}
& Symbol Detection &   Supervised    & RNN \cite{mosleh2017brain}, DNN \cite{samuel2017deep} \\
\cline{2-4}
& Channel Coding & Supervised  &   DNN \cite{ gruber2017deep}, RNN \cite{kim2018communication}\\
\cline{2-4}
& End-to-end learning & Supervised & Autoencoder \cite{o2017introduction}\\
\cline{2-4}
& Dynamic Spectrum Access & Reinforcement learning & DRL \cite{Chang_DSA_19}\\
\cline{2-4}
\hline

\multirow{4}{*}{\centering Network Layer} & Fault recovery and analysis  & Unsupervised &   Self-organizing maps \cite{gomez2015automatic}\\
\cline{2-4}
& Energy Optimization &   Supervised/unsupervised    & DNN\\
\cline{2-4}
& Resource management and scheduling & Reinforcement learning  &   DRL \cite{chinchali2018cellular}\\
\cline{2-4}
& Cell-sectorization & Reinforcement learning & DRL \cite{shafin2019self}\\
\hline
\end{tabular}
\label{table:PHY_and_Network_Layer}
\end{table*}

\subsection{Research Thrusts in AI for Wireless}
From a point-to-point link's perspective, \cite{o2017introduction} demonstrated that an end-to-end trained deep neural network (DNN)-based system performs identically to, and (under certain cases) outperforms, a conventional communication system. 
Deep learning has also been used for devising computationally efficient approaches for physical (PHY) layer communication receiver modules. 
Under the umbrella of supervised learning, \cite{samuel2017deep} presents a deep learning framework, called DetNET, for MIMO symbol detection. 
DetNET has been able to achieve near optimal detection performance, while providing $30$ times faster real-time implementation compared to a semi-definite relaxation-based approach. 
A recurrent neural network (RNN)-based detection strategy using supervised learning is introduced in \cite{mosleh2017brain} for MIMO-OFDM systems, and is shown to outperform traditional detection techniques under channel non-linearity. Convolutional neural network (CNN)-based supervised learning techniques can also be utilized for channel estimation problems offering better generalization ability and robustness to channel distortions \cite{neumann2018learning}. Another PHY layer application for supervised learning is channel decoding, where deep learning based decoding solutions have shown potential for classical codes such as convolutional and Turbo codes \cite{kim2018communication}, as well as for rather recent Polar codes \cite{gruber2017deep}. 
Besides point-to-point links, deep learning approaches have also been applied to wireless network design. 
Using an unsupervised learning framework, \cite{gomez2015automatic} develops an automatic fault detection and root cause analysis technique for LTE networks based on self-organizing maps. 
Meanwhile, deep reinforcement learning (DRL) has been applied for designing efficient spectrum access~\cite{Chang_DSA_19} and scheduling strategies~\cite{chinchali2018cellular} for cellular networks. 
Automatic cell-sectorization for cellular network coverage maximization is another area where DRL has shown tremendous potential \cite{shafin2019self}.
\subsection{Industry and Standardization} 
\begin{figure}[!th]\centering
\includegraphics[scale=0.3]{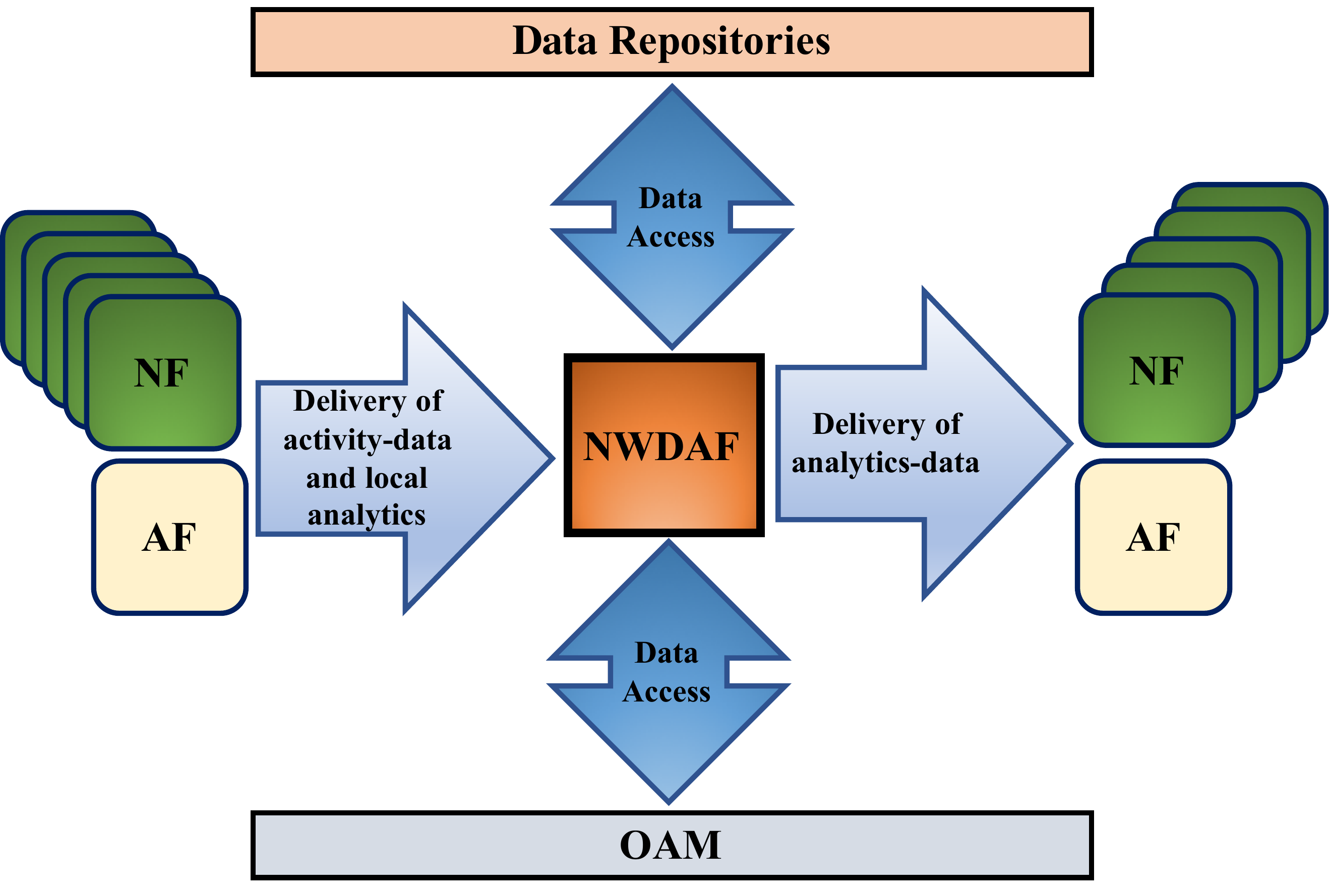}
\caption{3GPP 5G network automation}
\label{fig:3GPPautomation}
\end{figure}

Standards bodies have taken the first steps towards providing a framework for integrating AI models within planning, operation, and healing of future cellular networks. 
The third generation partnership project (3GPP) has defined a so-called network data analytics function (NWDAF) specification for data collection and analytics (including AI) in automated cellular networks~\cite{3GPP_NWDAF}. 
The standardization specifies only the interfaces to the NWDAF block, as shown in Fig.~\ref{fig:3GPPautomation}. By leaving the AI model development to implementation, 3GPP provides adequate flexibility for network vendors to deploy AI-enabled use cases. The inbound interfaces ingest data from various sources such as operation, administration, and maintenance (OAM), network function (NF), application function (AF), and data repositories, while the outbound interfaces relay the algorithmic decisions to the NF and AF blocks, respectively. 

\begin{figure}[!th]\centering
\includegraphics[scale=0.6]{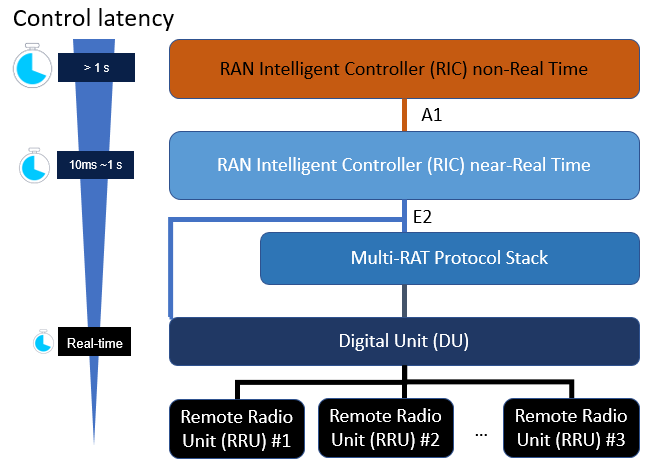}
\caption{O-RAN Architecture}
\label{fig:ORAN}
\end{figure}

In addition to 3GPP, five MNOs (AT\&T, China Mobile, Deutsche Telekom, NTT DOCOMO, and Orange) established the O-RAN Alliance in 2018, with the vision of an open and efficient radio access network (RAN) to leverage AI for automating different network functions and reduce operating expenses.
As of now, 21 MNOs and 81 network vendors including Samsung, Ericsson, Nokia, and ZTE are members of the alliance. 
The O-RAN architecture is shown in Fig.~\ref{fig:ORAN}, which includes AI-enabled RAN Intelligent Controller (RIC) for both non-real time (non-RT) and near-real time (near-RT), multi-radio access technology protocol stack. The non-RT functions include service and policy management, higher layer procedure optimization and model-training for the near-RT RAN functionality~\cite{O-RANwhitepaper}. The near-RT RIC is compatible with legacy radio resource management and enhances challenging operational functions such as seamless handover control, Quality of Service (QoS) management and connectivity management with AI. 
The O-RAN alliance has set up two work groups standardizing the A1 interface (between non-RT RIC and near-RT RIC) and E2 interface (between near-RT RIC and digital unit (DU) stack). 
The European Telecommunication Standards Institute has also initiated an Industry Specification Group on Experiential Networked Intelligence to define a cognitive network management architecture utilizing AI and context-aware policies to improve operator experience \cite{ENI_ETSI}.



\section{AI-Enabled Cellular Networks} 
\subsection{AI for PHY \& MAC Layers}
The PHY \& medium access control (MAC) layers are foundational layers of cellular networks where many technical innovations for 3G and 4G have taken place.  Below paragraphs discuss  use-cases where applying AI can potentially deliver improved performances within these layers.
\begin{asparadesc}
\item[Channel Estimation and Prediction.] 
Accurate channel state information (CSI) at the BS is critical for MIMO operation.
In massive MIMO systems, allocating pilot signals to derive complete CSI becomes prohibitive from the control overhead perspective. 
To reduce the pilot overhead, existing 5G NR standards limit the number of pilot signals to be significantly smaller than the number of antenna ports.
In this case, learning-based approaches can be adopted for tackling this channel estimation problem. 
It is shown in \cite{neumann2018learning} that MMSE channel estimators can be learned with low complexity using DNNs, and the learned estimator is shown to be optimal for some idealized channels.
\end{asparadesc}


\begin{asparadesc}
\item[Receive Processing.]
MIMO symbol detection constitutes a key module within the signal processing chain of communication receivers. For example, assuming the availability of receiver CSI, the optimal strategy is to apply the maximum likelihood detector. However, their performance is quite sensitive to model inaccuracies and/or CSI estimation errors. On the other hand, learning-based approaches can provide robust performance without relying on detailed channel models.  For example, the works in~\cite{samuel2017deep, mosleh2017brain} show that through end-to-end training of DNNs, AI models can outperform conventional MIMO symbol detection approaches even under imperfect receiver CSI.
Meanwhile, AI models can also be applied for interference cancellation to improve receiver performance.

\end{asparadesc}

\begin{asparadesc}
\item[Channel Decoding.]
AI approaches can be used for channel decoding in either an integrated or a stand-alone manner. In the first case, DNNs are utilized in conjunction with conventional approaches for obtaining performance gains.
For instance, as a variant of belief propagation decoding, the weights of the tanner graph can be learned using a DNN.
These schemes are particularly suitable for long block-length codes where learning the underlying structure of encoded blocks requires exorbitant amount of training and entails significant complexity. 
On the other hand, stand-alone DNN-based strategies are able to perform close to maximal aposteriori probability decoding for short block length communications~\cite{gruber2017deep, kim2018communication}. 
\end{asparadesc}

\begin{asparadesc}
\item[Random Access \& Dynamic Spectrum Access.] Spectrum access will be a critical problem for Beyond-5G and 6G networks. 
Existing methods mainly focus on designing spectrum access protocols under specific models so that efficient solutions can be achieved. Due to the heterogeneous nature of future cellular networks, such model-dependent solutions can not effectively adapt to real environment. 
Learning-based random access and dynamic spectrum access (DSA) strategies can be deployed in a distributed fashion to support spectrum access of massive number of devices. 
A DRL-based distributed DSA strategy is introduced in~\cite{Chang_DSA_19} showing devices could learn near-optimal spectrum access strategies without prior knowledge of the underlying network statistics.
\end{asparadesc}

\begin{figure*}[h!]
\centering
\includegraphics[width=0.9\linewidth]{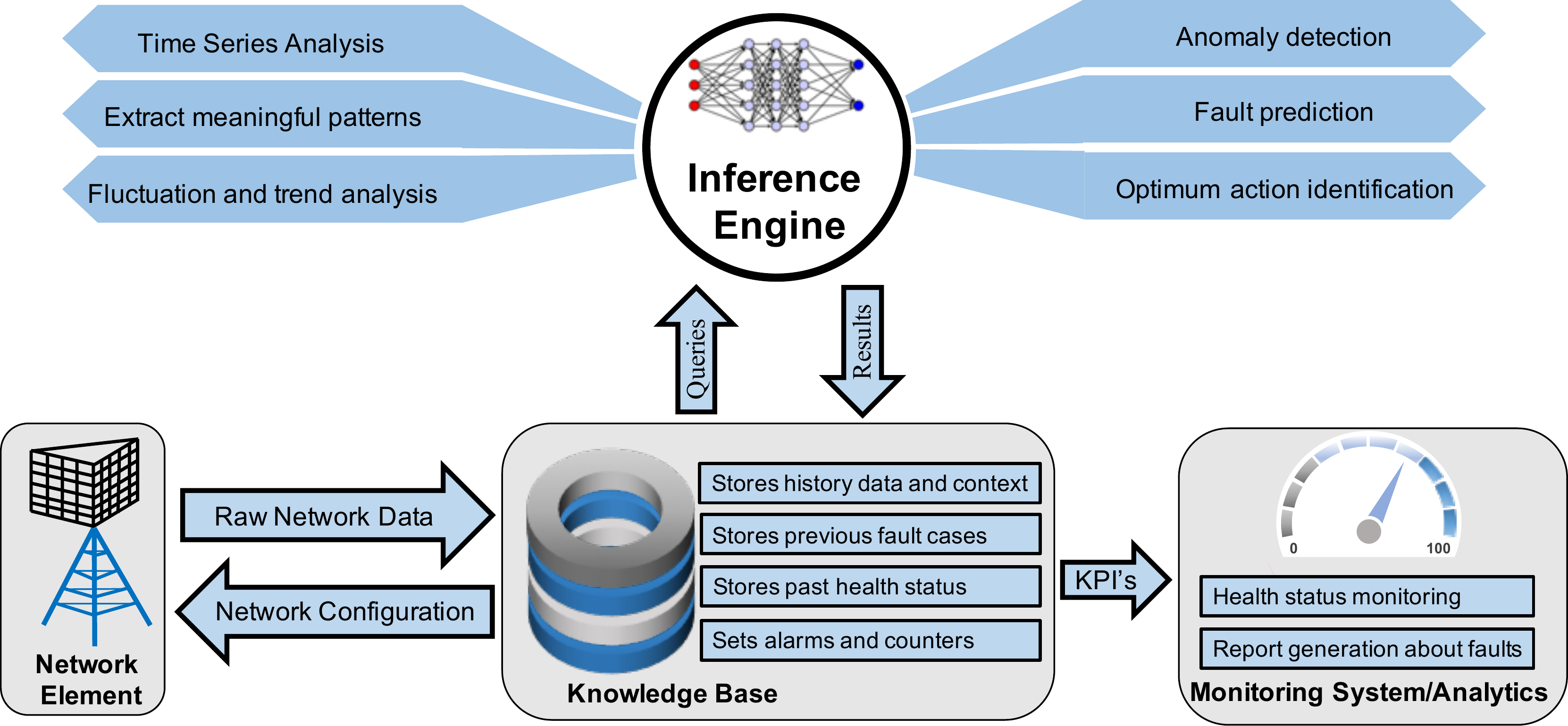}
\captionsetup{margin= {20pt},justification=centerlast,skip=-5pt,font=normalsize}
\vspace{0.3cm}
\caption[width=0.9\linewidth]{{AI-enabled fault identification and self-healing system.}}
\captionsetup{justification=centering}
\label{fig:fault_analysis}
\end{figure*}

\subsection{AI for the Network Layer}
The unrelenting demand for mobile data traffic imposes significant operational challenges for MNOs. 
The dense cell deployment for 5G will create increased network complexity requiring MNOs to devote additional resources for planning, operation, and trouble-shooting their 5G networks. 
As shown in Fig.~\ref{fig:fault_analysis}, an AI-enabled fault identification and self-healing system, within the framework of Self-Organizing Network, can be introduced so that MNOs can reduce their OPEX, reduce recovery time, and provide improved service quality to their end consumers. 
The following paragraphs discuss use-cases for each of the above aspects. 

\begin{asparadesc}
\item[Fault recovery (Root Cause Analysis, RCA).]
Each BS provides various data sources designated as key performance indicators (KPIs) to an operations support system. 
These KPIs typically consist of performance management (PM) counters sent periodically (typically every 15 minutes). 
The PM data reflect the state and behavior of the system. 
A subset of these data provide aggregated metrics reflecting the level of service accessibility, service retainability, service availability, service quality, and service mobility.
Troubleshooting is triggered in response to detecting one or more service quality anomalies. 
Manual troubleshooting requires human domain experts engaging in each RCA step including problem detection, diagnosis, and problem recovery. Since each BS reports thousands of KPIs during a single reporting interval, troubleshooting by a human expert, which is prevalent in most current networks, is non-trivial.

An AI-driven fault recovery system consists of two components, namely a knowledge base and an inference engine. The knowledge base consists of pre-processed historical data, derived using human domain expertise in combination with exploratory data analysis. The inference engine consists of an AI model or a set of rules applying the knowledge base data for RCA. Once the Inference Engine is trained, it can process real-time KPI data, detect anomalies and their associated root causes, and take remedial actions. For example, prior works \cite{DNA_HW, gomez2015automatic} have employed association rule-mining and self-organizing maps for detecting network anomalies and their underlying root causes.
\end{asparadesc}

\begin{asparadesc}
\item [Operation (AI-based Energy Optimization).]
Network function virtualization (NFV) will be an integral part of managing 5G networks. 
Using NFV, different virtual networks can be established in the same infrastructure providing diverse network services. 
Different virtual network functions (VNF) are created in different virtual machines, and VNF instances can be started, modified, or terminated on demand using a network management and orchestration system. Through container migration technologies, different VNF instances and the services provided by the VNFs can be shifted from one server to another. Usually data centers host the servers and are major source of power consumption in the network. 
By efficiently running the services in different servers, it is now possible to turn off a few servers, thereby saving power and OPEX. 

Given the large size of data centers and the complicated inter-connections, it is difficult to optimize their energy consumption in an error-free manner. An AI-managed data center can take into account a diverse set of network parameters and KPIs for optimizing the on-off operation of servers while ensuring uninterrupted services for the clients. Using the historical data collected by data center servers, it is possible to learn the pattern for usage and services. The collected data can also include information on resources such as CPU, storage, and network usage required for supporting each service. 


\end{asparadesc}

\begin{asparadesc}
\item [Operation (Scheduling).]
Scheduling plays a vital role in the operation of cellular networks. 
Due to the large number of control variables, to ensure manageable complexity, practical schedulers often implement simple metrics (e.g. rate proportional fair based), which are inherently sub-optimal. With the advent of newer 5G use-cases such as massive machine type communication (e.g. for industrial internet of things (IoT)), cellular networks will not only have to serve human users but potentially also thousands of low-cost low-power devices and sensors. Such devices will have different traffic characteristics than regular human users. For example, a sensor could wake up, relay its measurements via the cellular networks, and go back to sleep. 
Given the heterogeneous nature of future cellular networks, AI can be employed in practical schedulers for predicting the traffic arrivals and the amount of radio resources to allocate. 
In this regard, DRL has shown tremendous potential in solving challenging online decision-making tasks. 
In DRL, an agent, through direct interaction with its environment, learns to take better decisions over time. Recently, DRL has been applied for user scheduling for cellular networks and shown to provide superior performances over conventional strategies~\cite{chinchali2018cellular}. 
\end{asparadesc}

\begin{asparadesc}
\item[Network Planning (Self-Sectorization).]
Along with user-specific MIMO operation, cellular networks also need to create sector-specific wide beams to enhance network coverage, or transmit control and access signals. 
Selecting good broadcast beam parameters, such as elevation and azimuth beam-widths and antenna tilt, is important to maximize the network coverage. 
Traditionally, these parameters are set based on drive-test results, and once set, the parameters are kept unchanged for a long period of time, often months or years. 
This setup cannot be updated according to the change in users' distribution or mobility patterns, and hence results in strictly suboptimal solutions. A DRL-based framework can be introduced to learn the best broadcast beams \cite{shafin2019self}, and automatically update the antenna weights based on the changes in user distributions maximizing network coverage.
\end{asparadesc}

\section{Challenges and Roadmap }
Even though AI shows great promise for cellular networks, significant challenges remain to be overcome . In this section, we list the key challenges and provide a roadmap for realizing the vision of AI-enabled cellular networks for beyond-5G and 6G.

\subsection{Top Five Challenges}
\begin{asparadesc}
\item[Training Issues.] We foresee system overhead and availability of training data as two key impediments relating to training AI models for cellular networks. From a PHY and MAC layer perspective, training a cellular AI model using over-the-air feedback --- to update layer weights based on the back-propagation algorithm --- is likely prohibitively expensive in terms of uplink control overhead. Reducing training overhead is, therefore, a critical issue for the viability of PHY/MAC layer based AI models. Second, the separation of information across network protocol layers makes it difficult to obtain labeled training data. For example, training an AI model residing within a base-station scheduler may be challenging if it requires access to application layer information (e.g. end-user streaming video resolution quality).
\end{asparadesc}

\begin{asparadesc}
\item[Lack of Bounding Performance.] Unlike some other fields, it is important for cellular networks to be able to predict the worst-case behavior. Under traditional model-based approaches, it is generally well understood what the system output distribution would be, in response to a certain input distribution. This allows the system designer to prepare for a certain worst case scenario, while providing a minimum acceptable QoS or performance guarantee. On the other hand, due to their non-linear characteristics, it may be hard or even infeasible for AI approaches --- however well they perform in live networks --- to provide any worst-case performance guarantee. For smoothly integrating AI into cellular networks, it is crucial to ensure a tolerable and graceful degradation in a worst-case scenario.
\end{asparadesc}

\begin{asparadesc}
\item[Lack of Explainability.]  AI tools are often treated as black boxes as it is hard to develop analytical models to either test their correctness, or explain their behaviors, in a simple manner. The lack of explainability is a potential stumbling block in scenarios where AI is applied for real-time decision making (e.g. for vehicle-to-vehicle communications). Historically, cellular networks and wireless standards have been designed based on a mixture of theoretical analysis, channel measurements, and human intuition and understanding. This approach has proved amenable for domain experts to resort to either theoretical analysis or computer simulations to validate communication system building blocks. It is desirable for AI models to have similar levels of explainability when designed for cellular networks.
\end{asparadesc}

\begin{asparadesc}
\item[Uncertainty in Generalization.] If a communication task is performed using an AI model, it is often unclear whether the dataset used for training the model is general enough to capture the distribution of inputs as encountered in reality. For example, if neural network-based symbol detector is trained under one modulation and coding scheme (MCS), it is unclear how the system would perform for a different MCS level. This is not desirable in cellular networks, where MCS levels are changing adaptively due to mobility and channel fading, and it is important to predict system behavior in different scenarios. This is particularly true in mission-critical services where it is important to safeguard for rare events. Even though the learning engine didn't see the data point during training, it should still be able to generalize to unseen cases.
\end{asparadesc}

\begin{asparadesc}
\item[Lack of Interoperability.] Interoperability plays a critical part in today's increasingly complex cellular networks and frees the customers from vendor lock-in. Any inconsistency among AI-modules from different vendors can potentially deteriorate overall network performance. For example, some actions (e.g. setting handover threshold) taken by an AI-based module from one vendor could counteract the actions taken by another network module (which may or may not be AI-based) from a second different vendor. This could lead to unwanted handover occurrences between the original BS and the neighboring BS causing increased signaling overhead. Last, because an AI-based cellular network could have complex dependencies, it may be hard to pin-point which vendor equipment/AI module is responsible in case of any KPI degradation.
\end{asparadesc}


\subsection{Technology Roadmap} 
In light of the preceding challenges, from a technology roadmap perspective, new training algorithms and neural network architectures should be investigated to reduce training complexity and amount of training needed for PHY/MAC layer applications. Furthermore, interpretable and explainable AI tools will be crucial for obtaining insights into their decision making process. To maximize their robustness and minimize uncertainty in generalization, a canonical requirement could involve comparing the AI model output against a well-understood theoretical performance bound (e.g. maximum likelihood). 

Standards bodies such as 3GPP will have to carefully evaluate the underlying specification impact of AI models (e.g., neural network weights). If an air interface design uses a DNN for a certain transmission scheme, the standardization will have to carefully evaluate the associated signalling overhead (e.g., control information feedback). 
With the advent of IoT devices in addition to smart phones, given their low power requirements for devices at the edge of the network, the training of AI models could be split between the edge and the cloud. The specification could consider newer learning use-cases based on federated learning, i.e. distributed learning, at edge devices \cite{DBLP:journals/corr/abs-1809-00343}. 

Lastly, it is inevitable that future cellular networks will devise newer application scenarios that utilize features not just based on air interface data, but could also utilize cross-layer (e.g. application layer) information, for operating their AI models. Possibly the models could even incorporate multiple sensory modalities (e.g. based on vision, smell, hearing etc.). From a network design perspective, to maximize ease of deployment, it is desirable to provide clean interfaces, both within and across protocol stack layers, for providing feature inputs to AI models.

\subsection{Deployment Roadmap}
Given the incipient nature of deploying AI for wireless applications and the high levels of service guarantees required by MNOs, it is obligatory to employ AI in a phased manner. This will facilitate system designers to apply their lessons during initial AI deployments, and subsequently refine their AI tools and testing methodology. A first consideration relates to the time-scales for deploying AI in cellular. It may be preferable initially for AI models to operate across longer time-scales (e.g. order of minutes or hours), so that human domain experts can override model recommendations, if needed. 

Fail-safe mechanisms are desirable for minimizing impact of cascading errors due to unforeseen AI model outputs. Consider an example where an AI-based scheduler adapts its resource allocation (e.g. MCS) based on the underlying radio environment. If the actions of the AI model result in an unacceptably large block error rate, the scheduler could override the model and re-initialize with the lowest MCS level for robust transmissions.

Additional robustness can be added if the AI model adapts its actions based on human expert feedback. One such scenario is where, upon detecting a network anomaly, an AI model outputs a certain root cause explanation that appears erroneous. If the expert can provide feedback regarding the incorrect decision, the model can refine and improve its decision-making until it reaches a point where its decisions are indistinguishable from an expert.

\section{Conclusion}
AI promises to revitalize wireless communications in the 21st century. 
This article has overviewed the state-of-the-art research topics, identified key obstacles, and presented a roadmap towards fulfilling the potential of AI in cellular networks. The formidable technological barriers should inspire fundamental research and engineering ingenuity in this field. We believe this is the surest path towards realizing the vision of AI for Beyond-5G and 6G cellular networks.

\bibliographystyle{IEEEtran}
\bibliography{IEEEabrv,MIMO_OFDM}
\end{document}